\documentclass[aps,prl,twocolumn,superscriptaddress,showpacs,floats,floatfix,preprintnumbers,amsmath,amssymb]{revtex4}
\usepackage{graphicx}
\usepackage{dcolumn}
\usepackage{bm}

\begin{document}

\preprint{(to be submitted to Phys. Rev. Lett.)}

\title{Structural Metastability of Endohedral Silicon Fullerenes}

\author{Alex Willand}
\affiliation{Department of Physics, Universit\"{a}t Basel, Klingelbergstr. 82, 4056 Basel, Switzerland}
\author{Matthias Gramzow}
\affiliation{Fritz-Haber-Institut der Max-Planck-Gesellschaft, Faradayweg 4-6, 14195 Berlin, Germany}
\author{S. Alireza Ghasemi}
\affiliation{Department of Physics, Universit\"{a}t Basel, Klingelbergstr. 82, 4056 Basel, Switzerland}
\author{Luigi Genovese}
\affiliation{ European Synchrotron Radiation Facility, 6 rue Horowitz, 38043 Grenoble, France}
%\affiliation{Commissariat \`a l'Energie Atomique, INAC/SP2M/L\_Sim, 17 rue des Martyrs, 38054 Grenoble cedex 9, France}
\author{Thierry Deutsch}
\affiliation{Commissariat \`a l'Energie Atomique, INAC/SP2M/L\_Sim, 17 rue des Martyrs, 38054 Grenoble, France}
%\affiliation{Commissariat \`a l'Energie Atomique, INAC/SP2M/L\_Sim, 17 rue des Martyrs, 38054 Grenoble cedex 9, France}
\author{Karsten Reuter}
\affiliation{Fritz-Haber-Institut der Max-Planck-Gesellschaft, Faradayweg 4-6, 14195 Berlin, Germany}
\affiliation{Fritz-Haber-Institut Dept. Chemie, Technische Universit\"{a}t M\"{u}nchen, Lichtenbergstr. 4,
85747 Garching, Germany }
\author{Stefan Goedecker}
\affiliation{Department of Physics, Universit\"{a}t Basel, Klingelbergstr. 82, 4056 Basel, Switzerland}

\date{\today}

\begin{abstract}
Endohedrally doped Si$_{20}$ fullerenes appear as appealing building blocks for nanoscale materials. We investigate their structural stability with an unbiased and systematic global geometry optimization method within density-functional theory. For a wide range of metal doping atoms, it was sufficient to explore the Born Oppenheimer surface for only a moderate number of local minima to find structures that clearly differ from the initial endohedral cages, but are considerably more favorable in terms of energy. Previously proposed structures are thus all metastable. 
\end{abstract}


\maketitle

As miniaturization techniques are reaching their ultimate limits, the interest in novel silicon based nanoscale devices increases. Notwithstanding, most common materials in nano sciences to date are carbon based such as fullerenes and nanotubes. Si based clusters and nanoparticles have also been studied extensively and it was shown that their electronic structure as well as their mechanical, optical and magnetic properties can be manipulated by changing their shape, size and composition. There is widespread hope that such Si nanomaterials may be basic building blocks for more complicated structures, such as wires and layers \cite{appel02,lieber02,zheng04,cui01,zhou03}. Unfortunately, up to now no stable Si building blocks have been found that are as chemically unreactive and symmetric, and therewith attractive for cluster assembled materials, as the carbon fullerenes. 

Endohedral doping with metal atoms is a primary avenue believed to stabilize cage-like Si geometries. In fact, clathrates are composed of corresponding polyhedral building blocks \cite{rachi05}. The exceptional elastic, thermoelectric, optoelectronic and super-conducting properties of these porous crystals \cite{sanmiguel99,tse00,gryko00,kawaji95} already illustrate the unique potential offered if novel materials could be tailored out of such Si-based subunits. Considering that C$_{20}$ forms the smallest known fullerene, Si$_{20}$ clusters represent a particularly interesting size in this context that should in principle be large enough to encapsulate a metal atom \cite{janssens07}. In contrast to the intrinsically unstable hollow Si$_{20}$ fullerene \cite{ho92}, endohedral doping with a range of metal atoms was indeed theoretically predicted to stabilize the cage structure \cite{sun02,kumar07}. 

In these, as well as in numerous equivalent theoretical studies on other cluster sizes, the stability was inferred from computed embedding and binding energies of relaxed structures. By construction, corresponding geometry optimizations lead, however, only to the next {\em local} minimum on the Born-Oppenheimer potential energy surface (PES). While a harmonic frequency analysis may ensure that this local minimum has indeed been reached, this still does not tell anything about the {\em global} PES features. In particular such an approach does not tell us if there are other energetically even more favorable minima, or if the present structure indeed corresponds to the global minimum. Starting the geometry optimization from several initial configurations \cite{hossain07,sporea07} or using several stages of symmetry constraints \cite{wang07,zdetsis07} may provide some information in this direction. Still, the corresponding exploration of the PES is by no means systematic, and the reliability of the deduced structural stability uncertain. 

In this work, we therefore reexamine the structure of metal-doped Si$_{20}$ clusters using a global and unbiased geometry optimization technique within density-functional theory (DFT). For essentially the entire range of previously proposed metal dopants this readily identifies significantly more stable structures that no longer correspond to endohedral fullerene cages. The latter configuration thus only corresponds to a {\em local} PES minimum, and the partial information we obtain on the surrounding barriers even suggests that this minimum is in most cases quite shallow. With a corresponding at best feeble metastability restricted to low temperatures, doped Si$_{20}$ fullerenes are unlikely useful building blocks for future nanoscale materials -- unless additional stabilization mechanisms are identified.

The minima hopping method (MHM) \cite{goedecker04} is designed to find the global minimum of complex polyatomic systems in an efficient way. The general idea is to limit repeated visits of the same local minima without penalizing crossings through important transition regions, such as hubs connecting superbasins of the potential landscape. The method is composed of an inner part, that attempts to escape from the current minimum by following short trajectories from molecular dynamics (MD), and an outer part that either accepts or rejects the new configuration by simple energy thresholding. A feedback mechanism on both parts allows to take advantage of the history of minima visited, as well as of the Bell-Evans-Polanyi principle, which correlates lower energy barriers with deeper basins \cite{roy08}.

In order to reach predictive quality, the PES explored by the MHM must be computed from first-principles. Here we use DFT as implemented in the BigDFT package \cite{genovese08} with valence-type pseudopotentials \cite{hartwigsen98} for this purpose. In the spirit of the dual MHM \cite{goedecker05} two levels of accuracy are considered to reduce the computational cost. During the MD escapes and in the initial stages of local geometry relaxations a coarser grid with smaller simulation boxes was chosen to define the employed adaptive wavelet basis. For the final geometry optimization and the evaluation of the total energy of the relaxed structure highly accurate parameter sets were used. 
In these calculations we rely on the widespread local-density approximation (LDA) \cite{goedecker96} as an efficient general-purpose approach to treat electronic exchange and correlation (xc). 
In order to check the accuracy of the LDA xc functional 
we recomputed the energetic order of the identified minima with gradient-corrected (PBE \cite{perdew96}) and hybrid (PBE0 \cite{ernzerhof99}, B3LYP \cite{stephens94}) functionals. The latter computations were done with the accurate all-electron full-potential code FHI-aims \cite{blum09} using the ``tier2'' basis-set composed of atomic-centered numeric orbitals. For the LDA and PBE functionals contained in both codes the obtained energetic differences agreed to within 150\,meV, thereby confirming the accuracy of the pseudopotentials employed in the initial BigDFT calculations and the near-completeness of the basis set used in the FHI-aims calculations.

\begin{table}
\begin{tabular}{|c|c|c|c|c|c|}
\hline              & preferred & \multicolumn{4}{c|}{energy gap to cage} \\
                    & structure & LDA  &PBE  &PBE0 &B3LYP \\
\hline Ba@Si$_{20}$ & broken    & 1.47 &1.75 &3.19 &0.70  \\
\hline Ca@Si$_{20}$ & broken    & 0.99 &0.98 &1.98 &0.72  \\
\hline Cr@Si$_{20}$ & Si$_{15}$ & 0.87 &1.07 &1.01 &0.39  \\
\hline Cu@Si$_{20}$ & Si$_{10}$ & 1.44 &1.83 &2.24 &2.12  \\
\hline K@Si$_{20}$  & broken    & 3.32 &3.72 &4.74 &2.77  \\
\hline Na@Si$_{20}$ & broken    & 1.36 &1.45 &1.57 &0.59  \\
\hline Pb@Si$_{20}$ & broken    & 2.63 &2.67 &2.98 &2.02  \\
\hline Rb@Si$_{20}$ & broken    & 2.13 &2.72 &3.52 &2.57  \\
\hline Sr@Si$_{20}$ & broken    & 1.39 &1.72 &2.31 &1.09  \\
\hline Ti@Si$_{20}$ & Si$_{16}$ & 0.68 &0.79 &1.15 &2.01  \\
\hline V@Si$_{20}$  & Si$_{16}$ & 1.52 &0.85 &1.42 &0.58  \\
\hline Zr@Si$_{20}$ & Si$_{16}$ & 1.43 &1.42 &3.02 &2.00  \\
\hline
\end{tabular} \\
\caption{ \label{tableI}
Results of the MHM search for the lowest-energy structure of doped Si$_{20}$ clusters, which either correspond to exohedral configurations with broken cages or smaller Si$_{n}$ cages with peripheral bud. Additionally shown is the energy difference (in eV) between this most stable structure encountered and the lowest-energy intact cage (see text). Summarized is the energetic data for a range of local, gradient-corrected and hybrid DFT xc functionals as obtained with FHI-aims \cite{blum09}. The listed energies are always 
the energies that correspond to the lowest energy spin state. } 
\end{table}

\begin{figure*} %The * figure goes over the full page width
\begin{center}
\setlength{\unitlength}{1cm}
\begin{picture}( 19.0,22.2) % figure dimensions
%first coloumn
\put(0.4,18.5){\includegraphics[width=3.7cm]{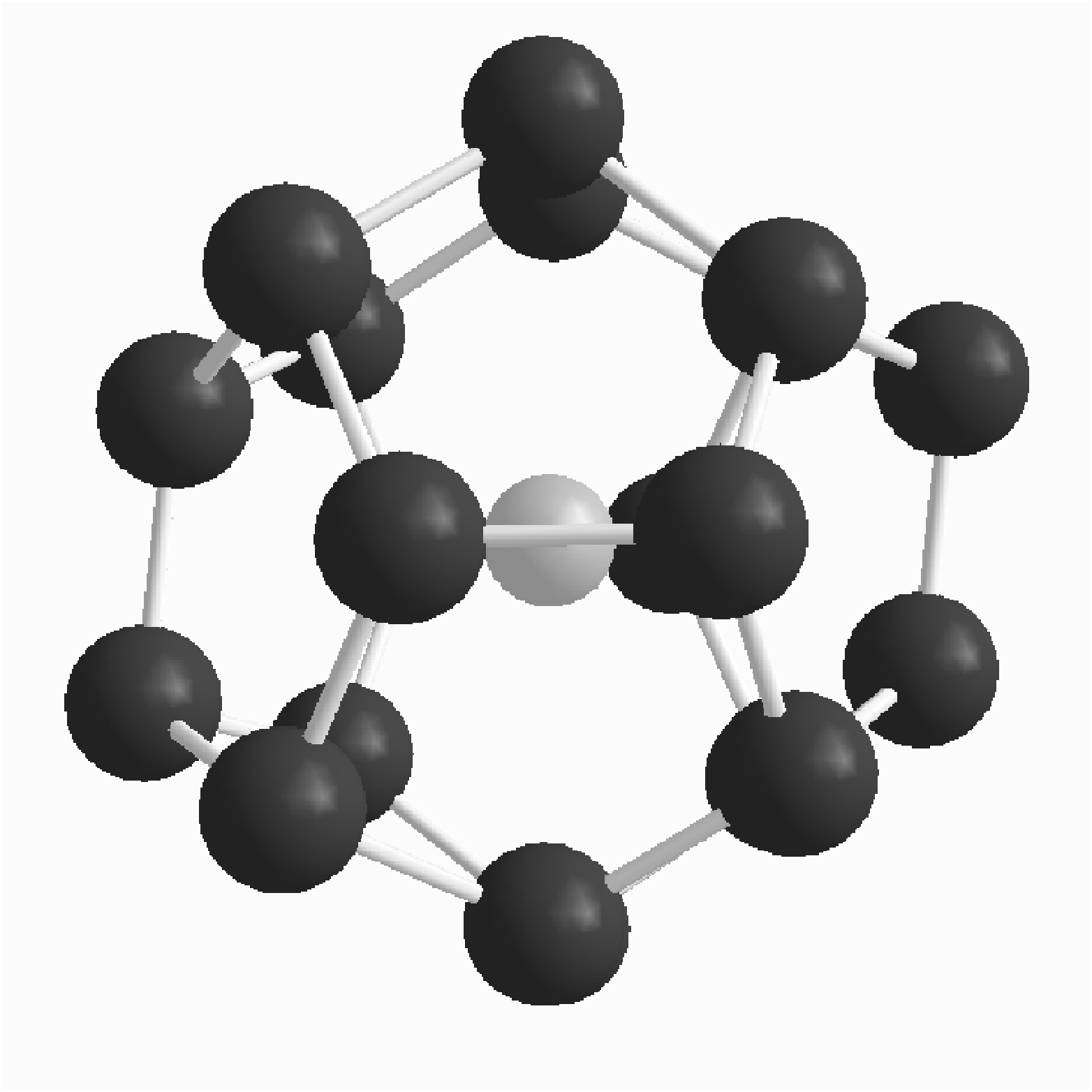}}
\put(0.4,14.8){\includegraphics[width=3.7cm]{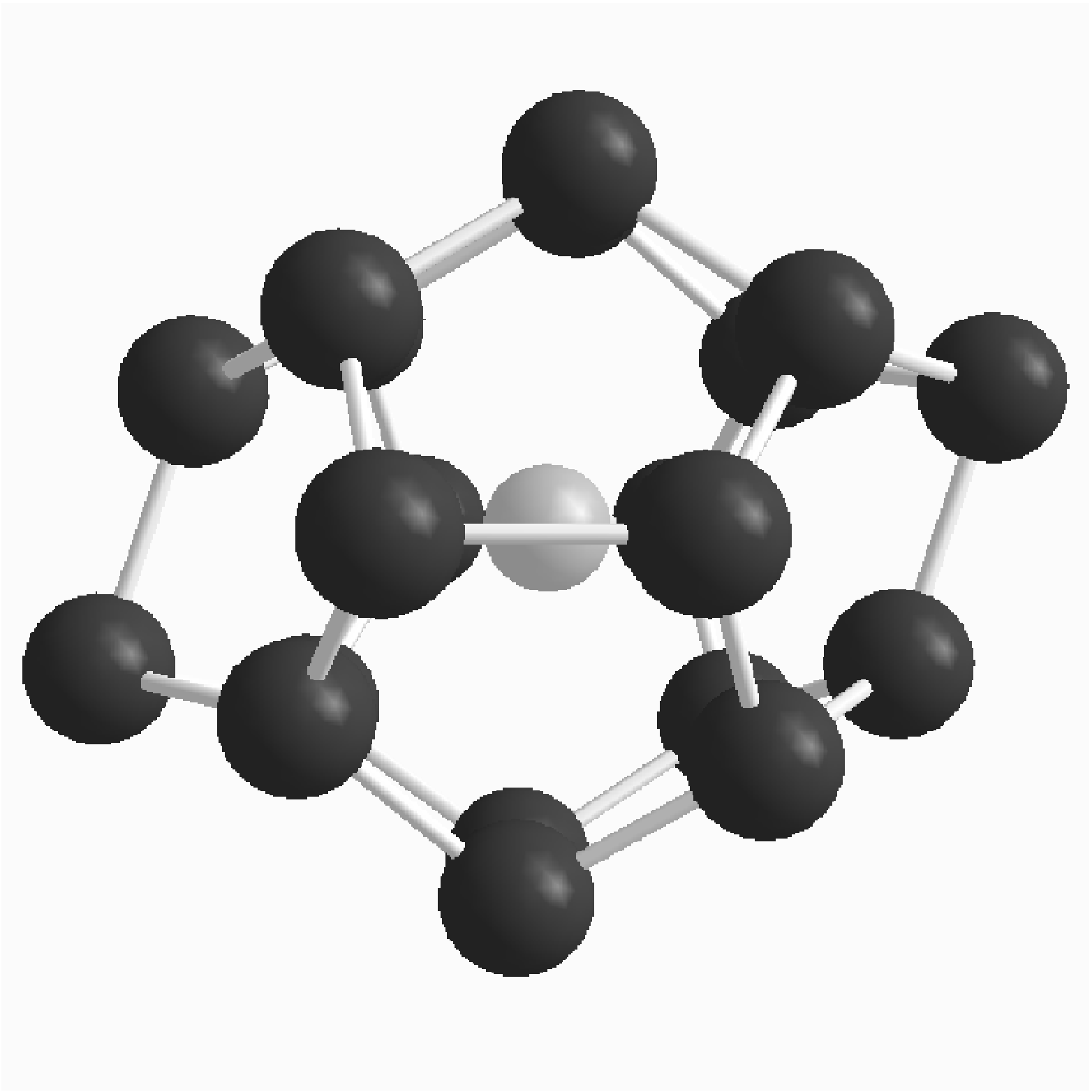}}
\put(0.4,11.1){\includegraphics[width=3.7cm]{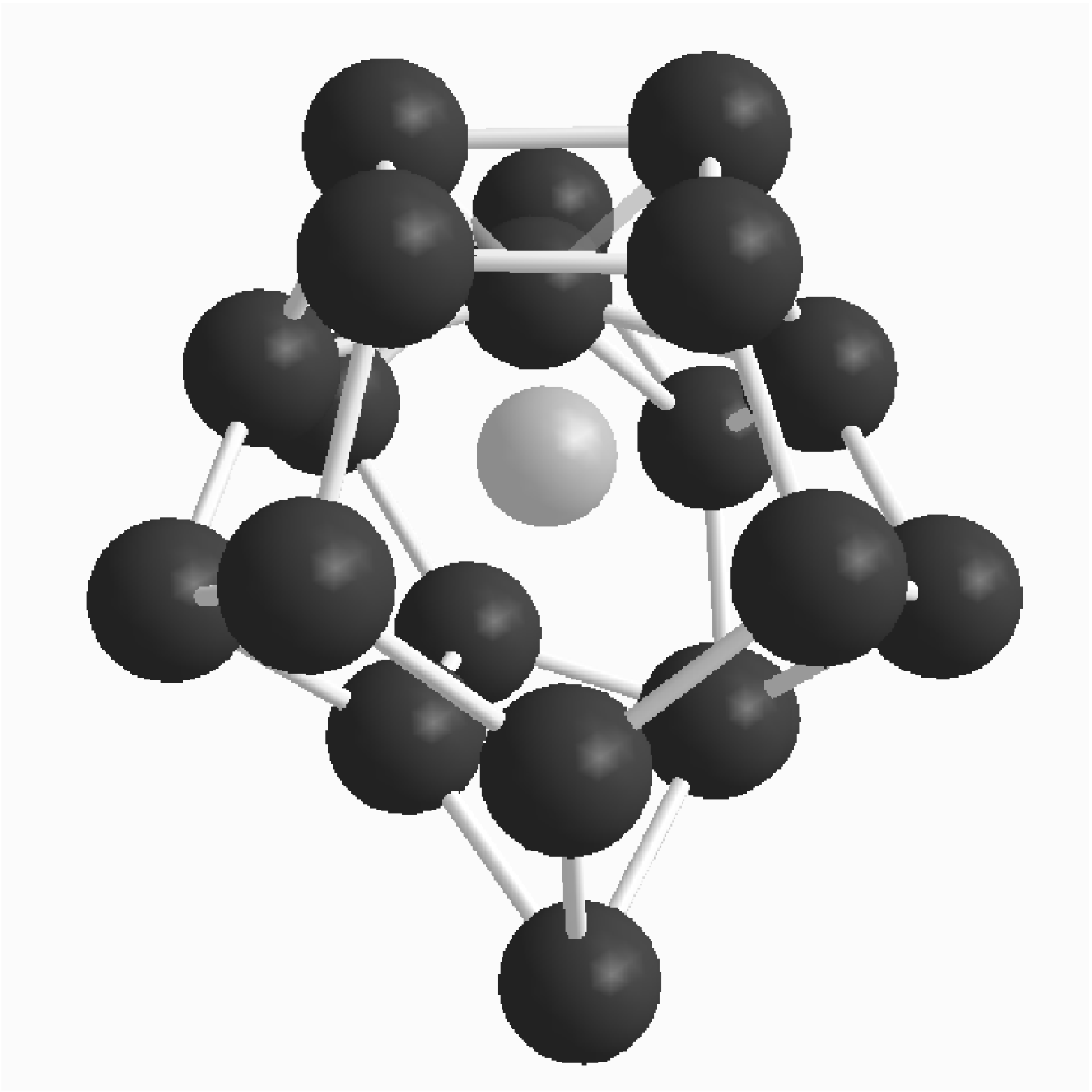}}
\put(0.4,07.4){\includegraphics[width=3.7cm]{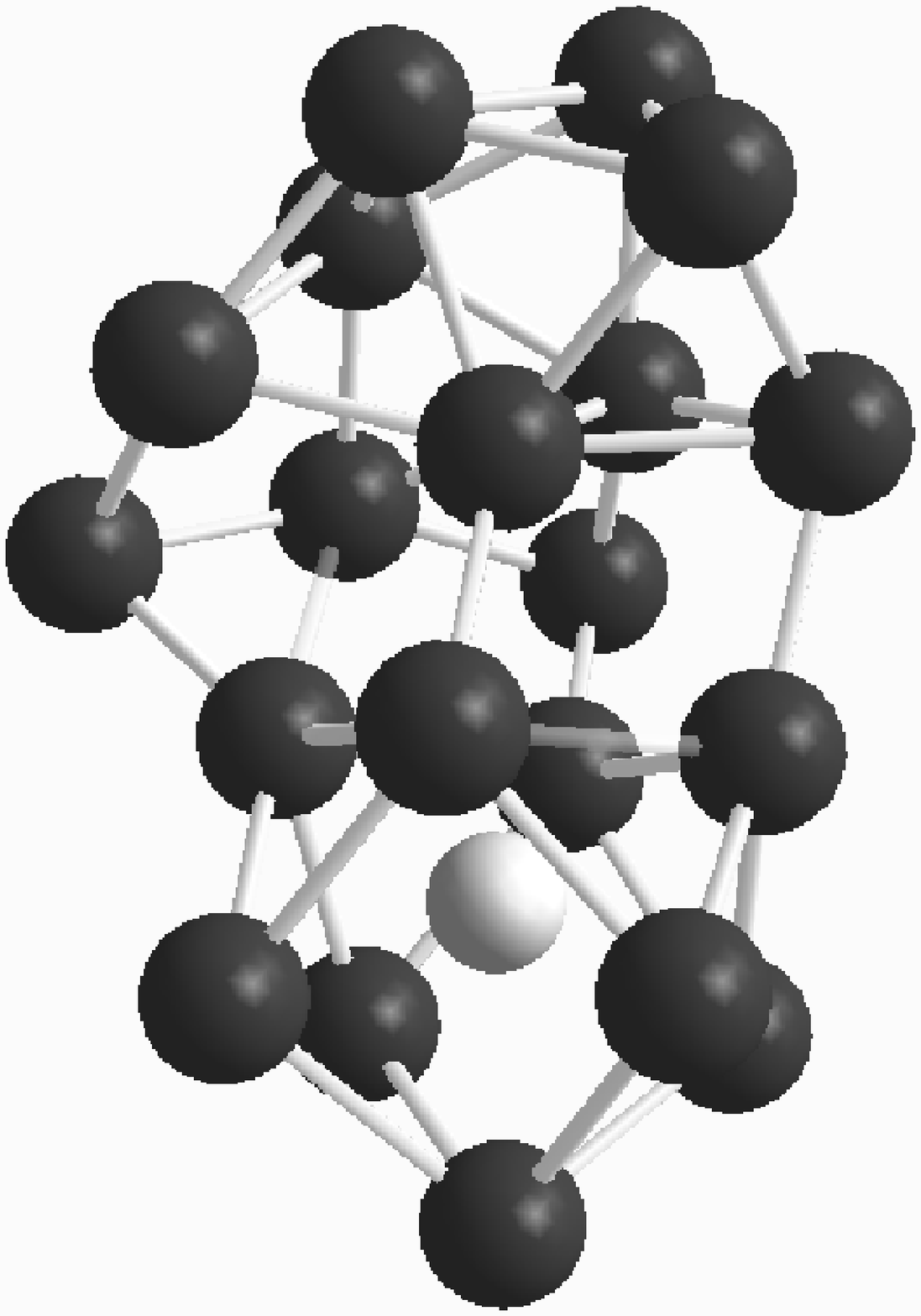}}
\put(0.4,03.7){\includegraphics[width=3.7cm]{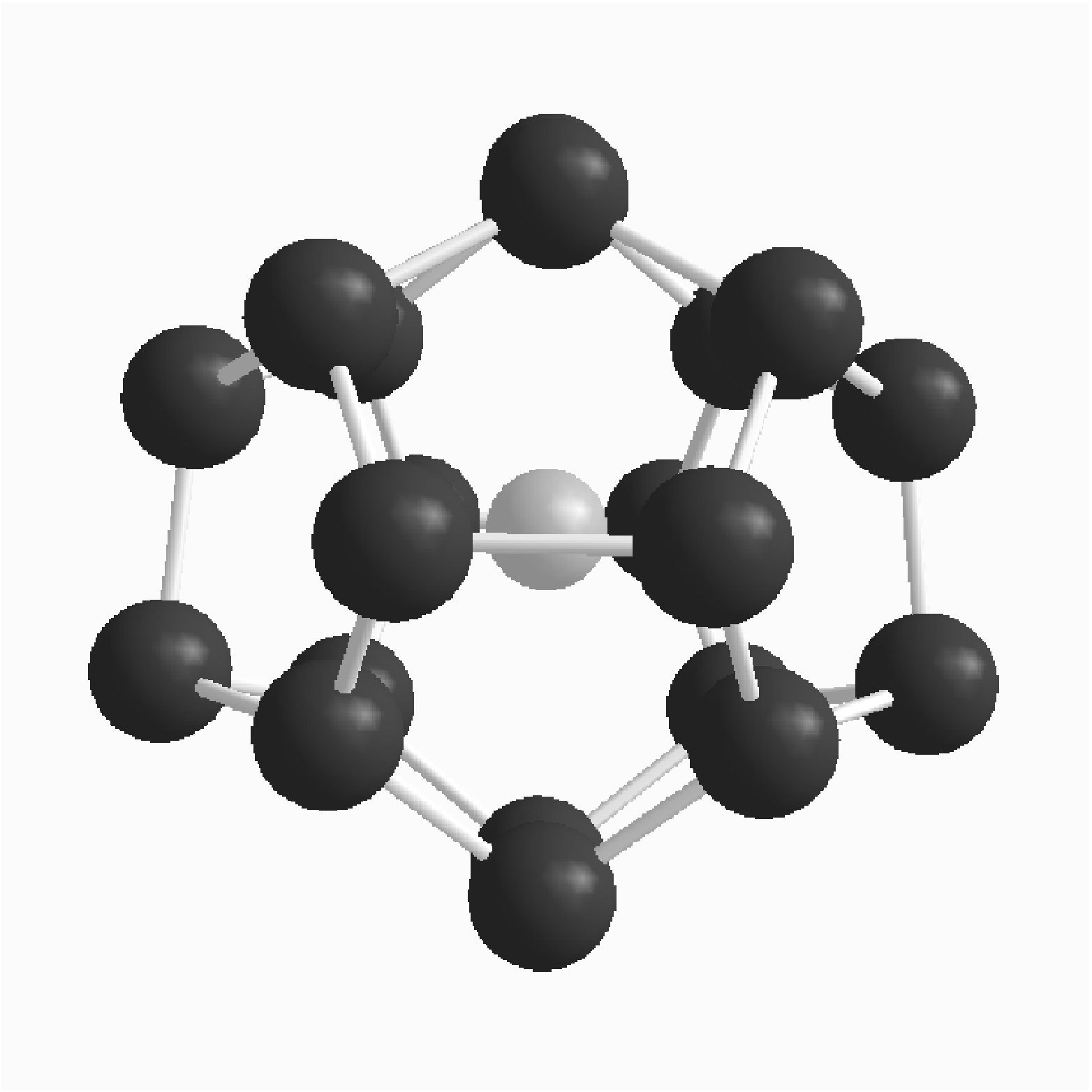}}
\put(0.4,00.0){\includegraphics[width=3.7cm]{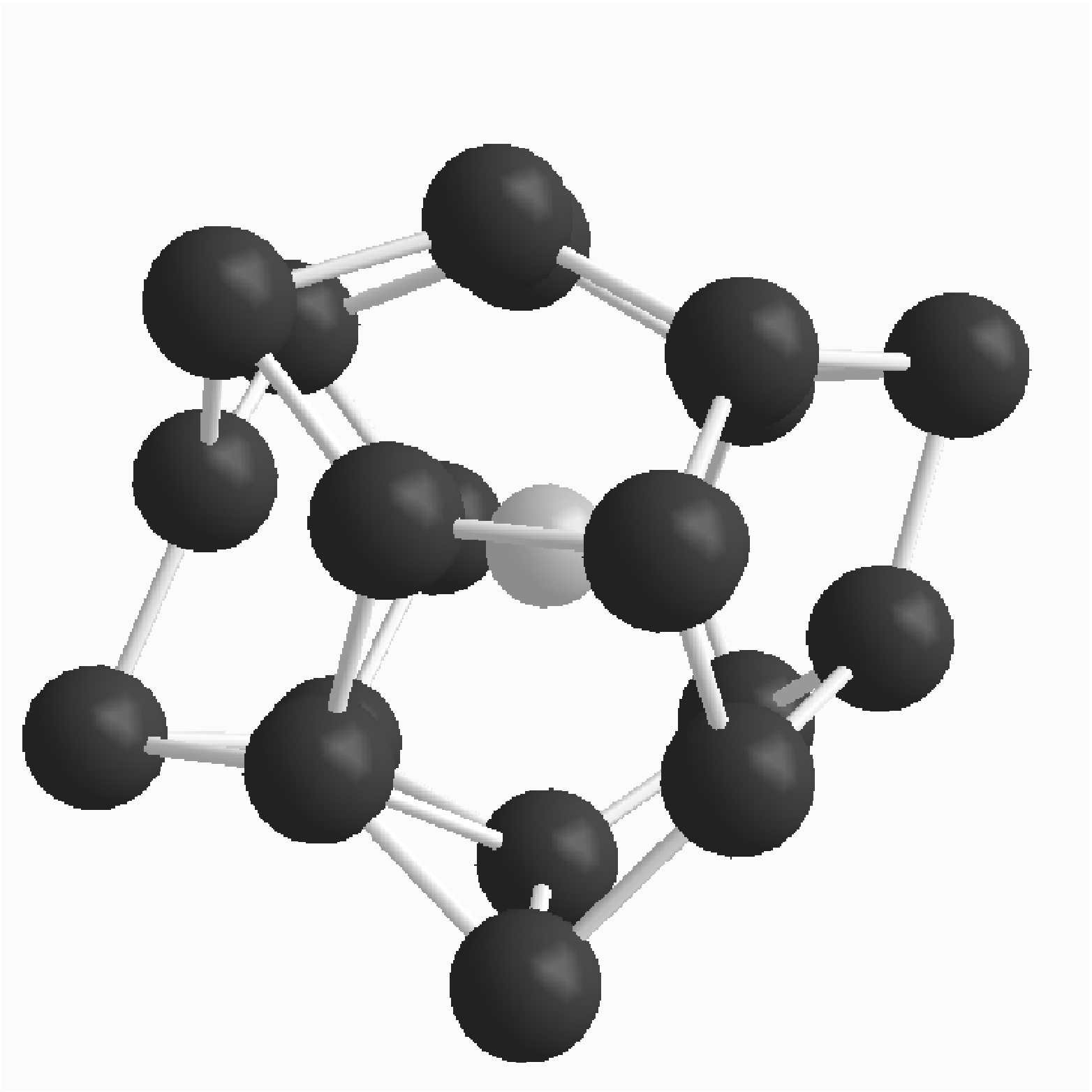}}
%second coloumn
\put(4.4,18.5){\includegraphics[width=3.7cm]{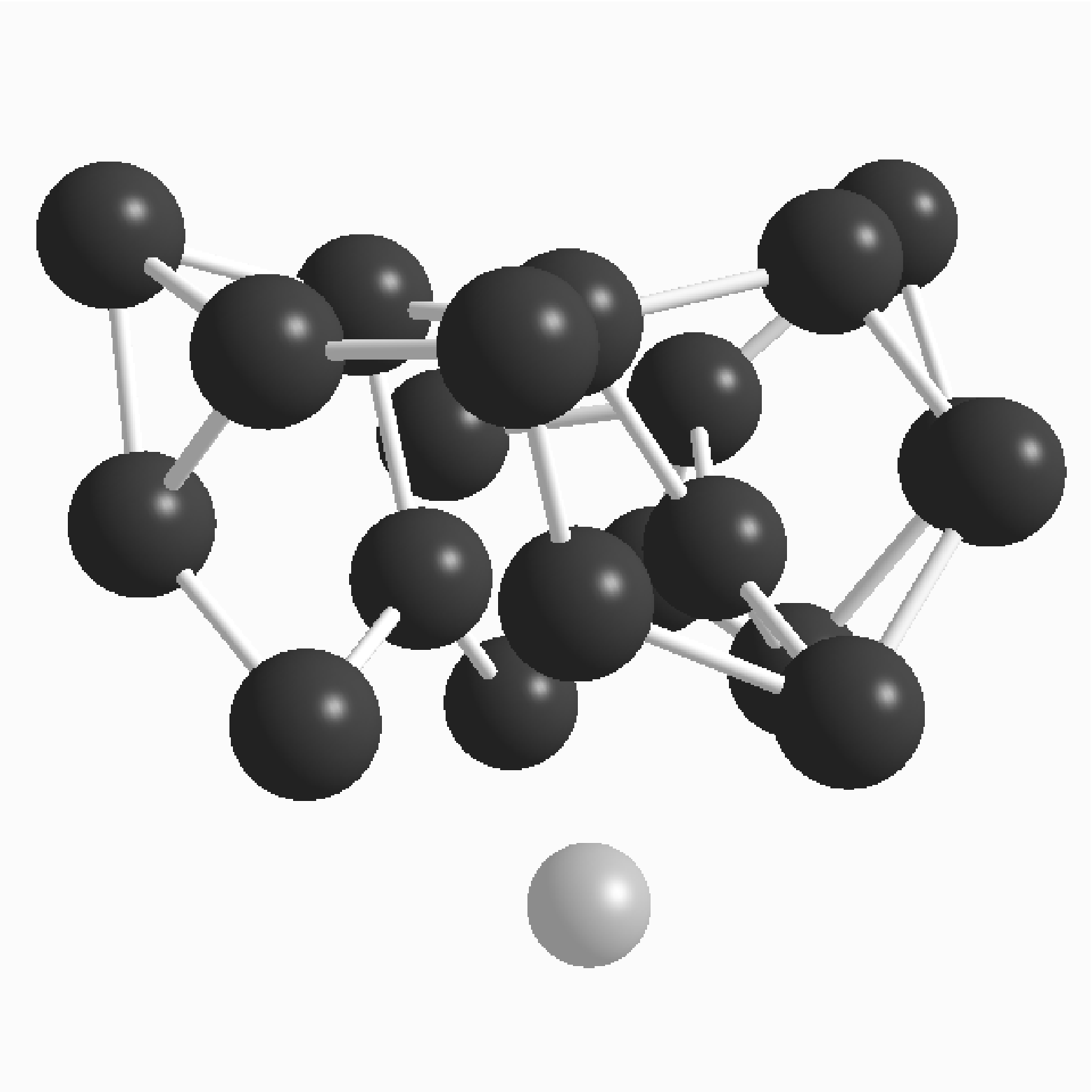}}
\put(4.4,14.8){\includegraphics[width=3.7cm]{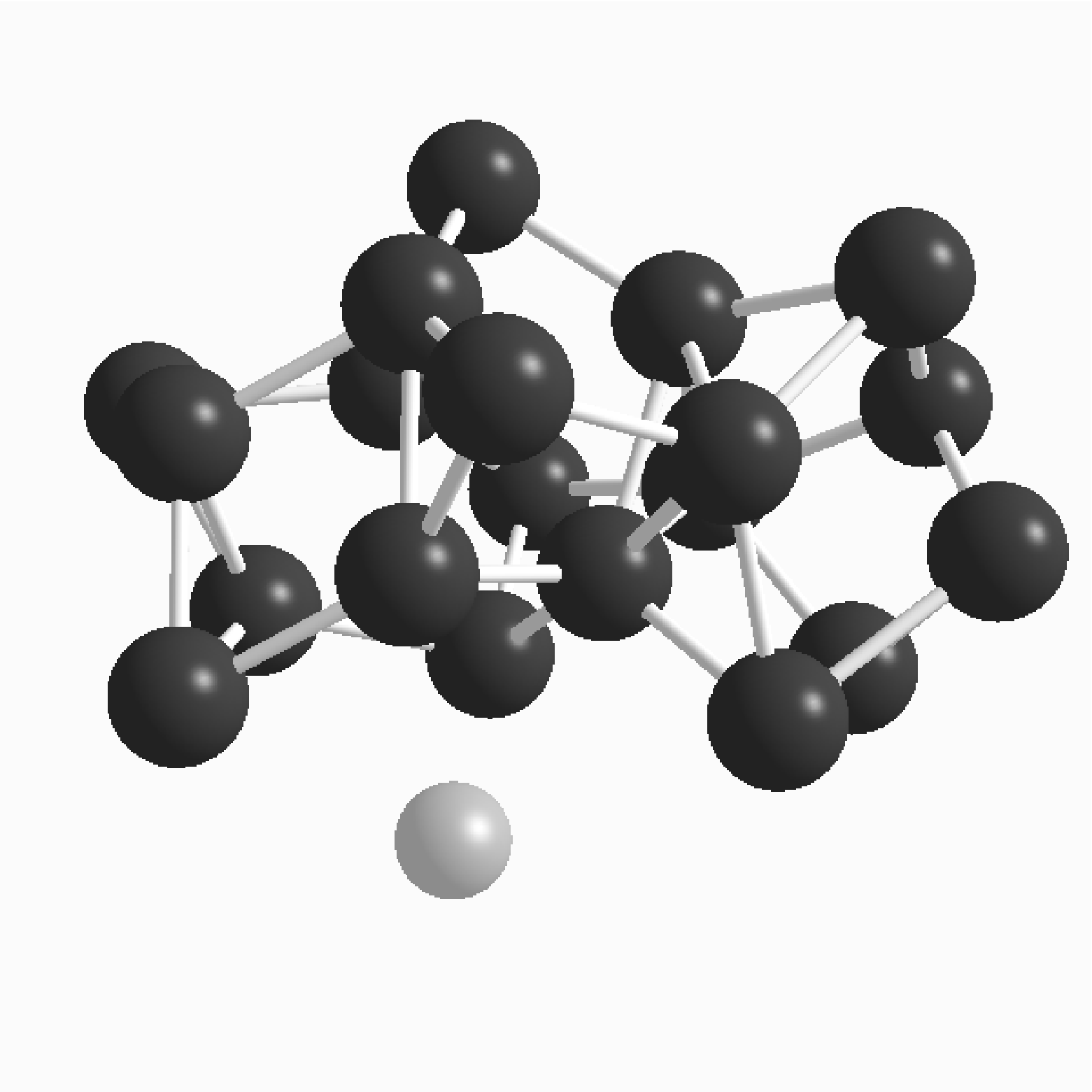}}
\put(4.4,11.1){\includegraphics[width=3.7cm]{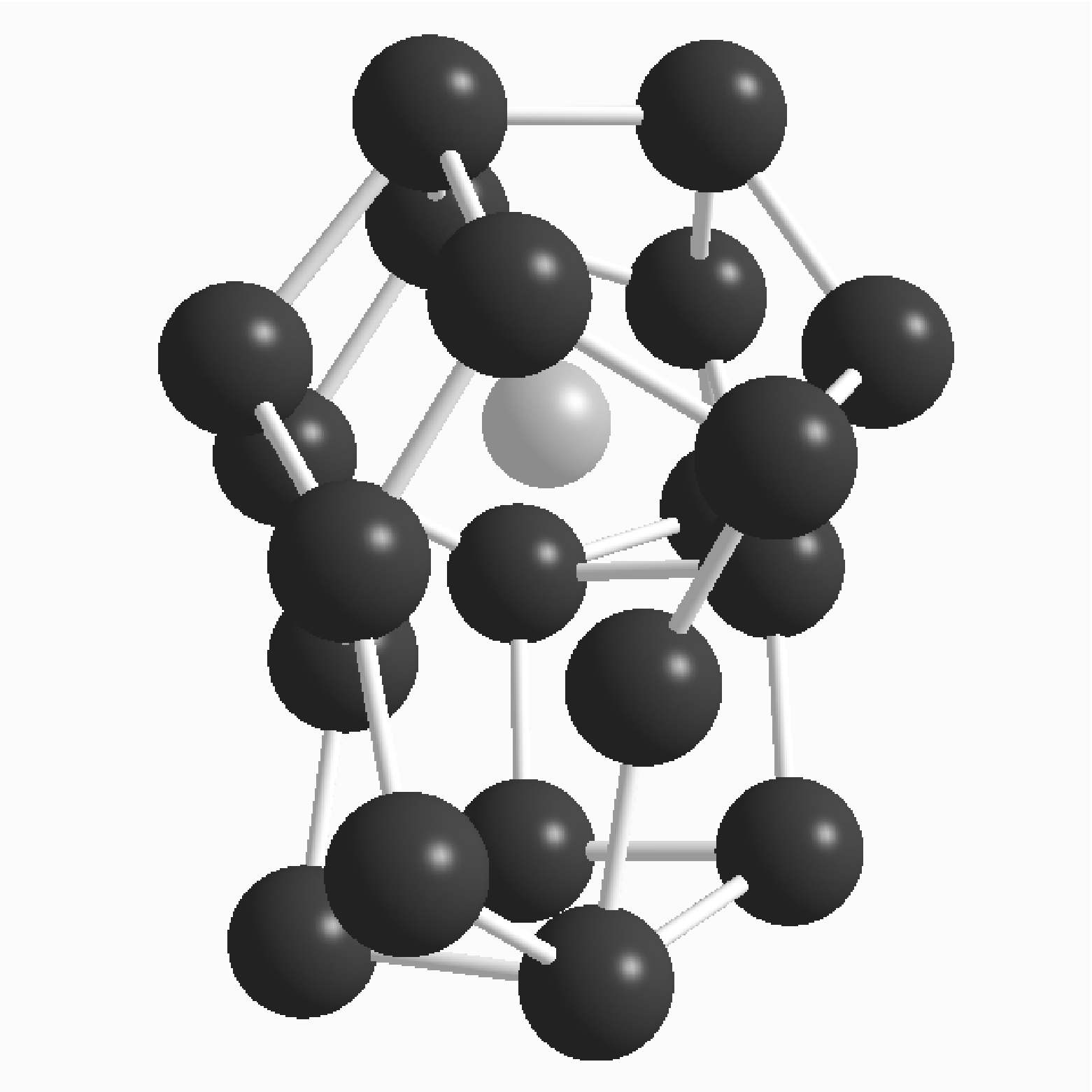}}
\put(4.4,07.4){\includegraphics[width=3.7cm]{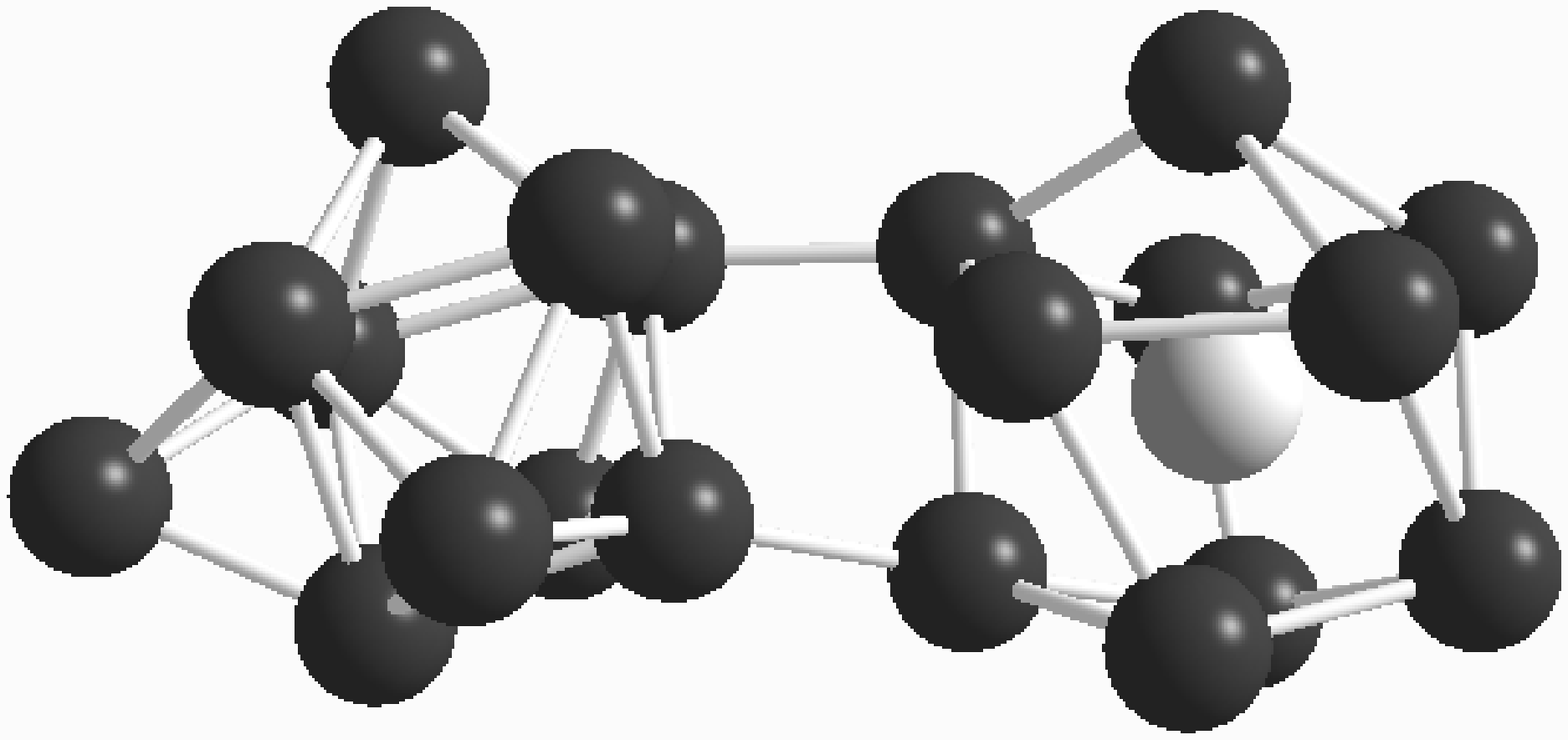}}
\put(4.4,03.7){\includegraphics[width=3.7cm]{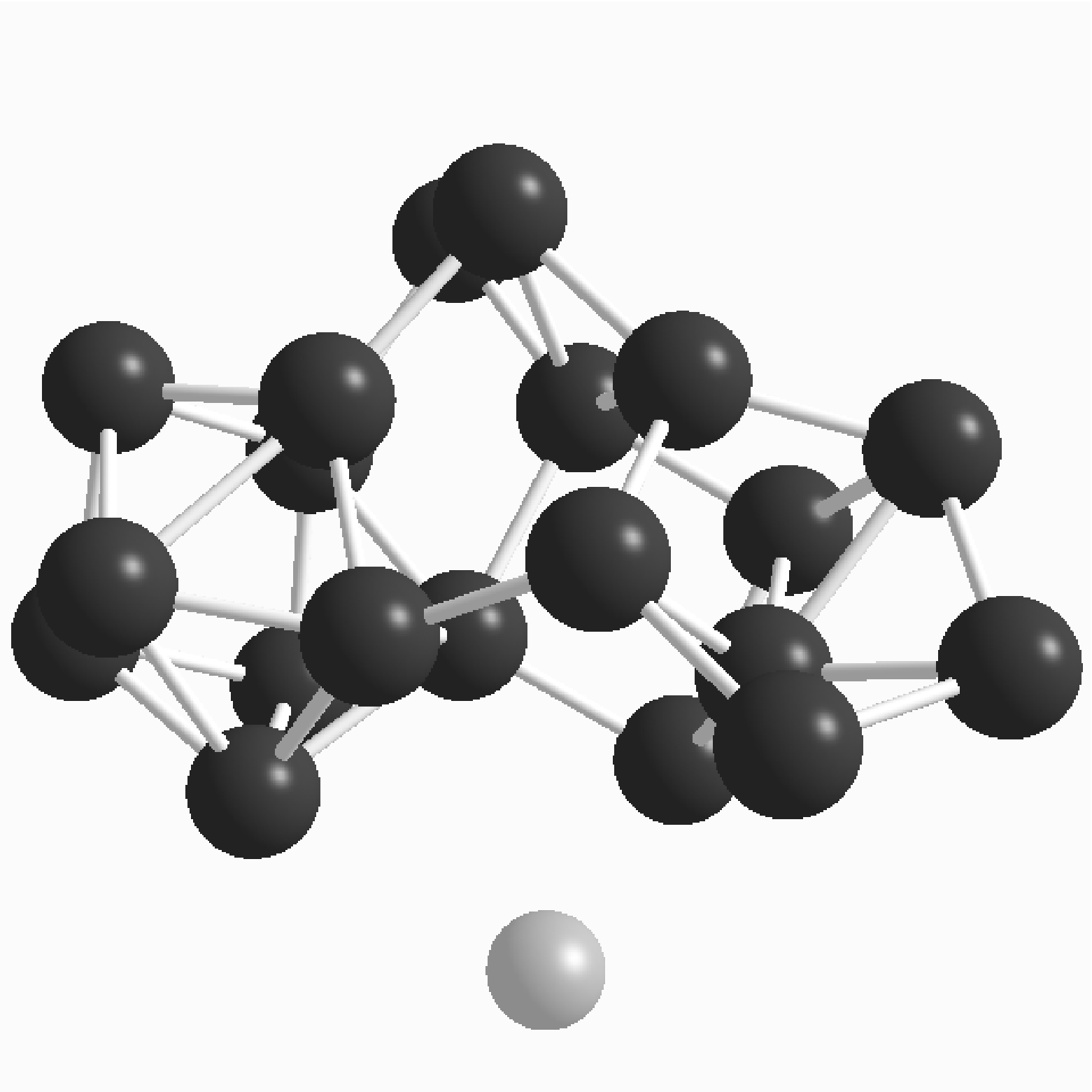}}
\put(4.4,00.0){\includegraphics[width=3.7cm]{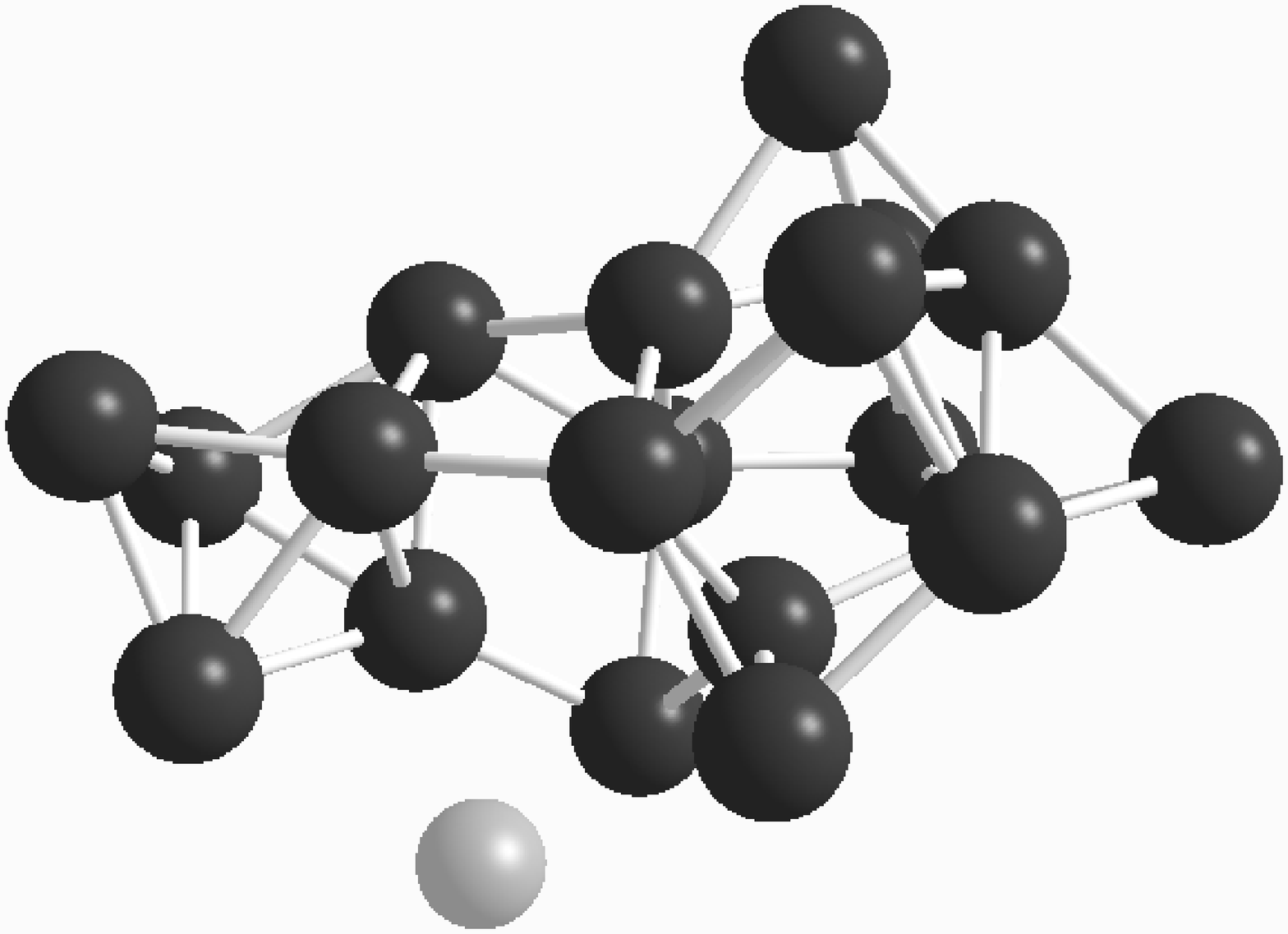}}
%third coloumn
\put(9.9,18.5){\includegraphics[width=3.7cm]{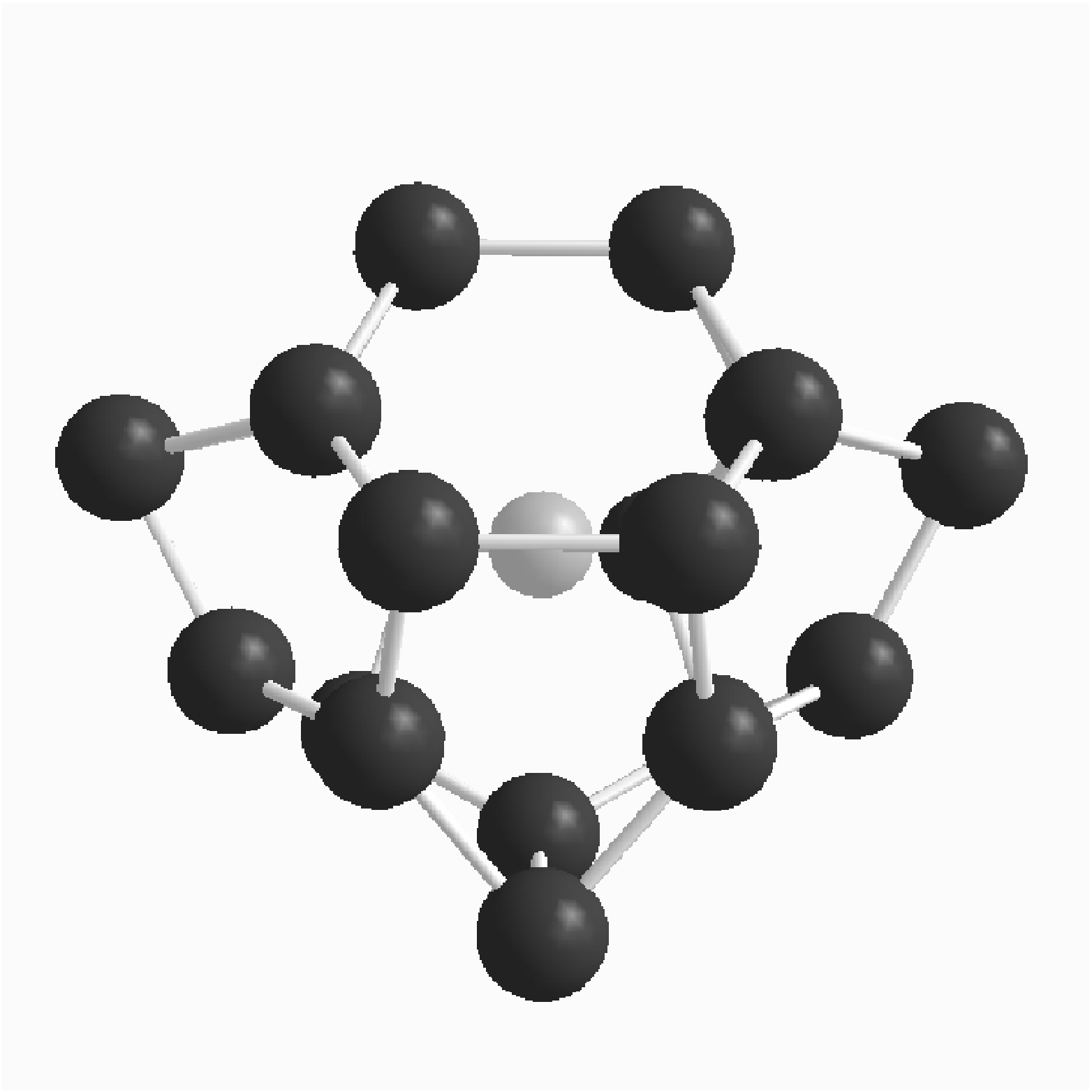}}
\put(9.9,14.8){\includegraphics[width=3.7cm]{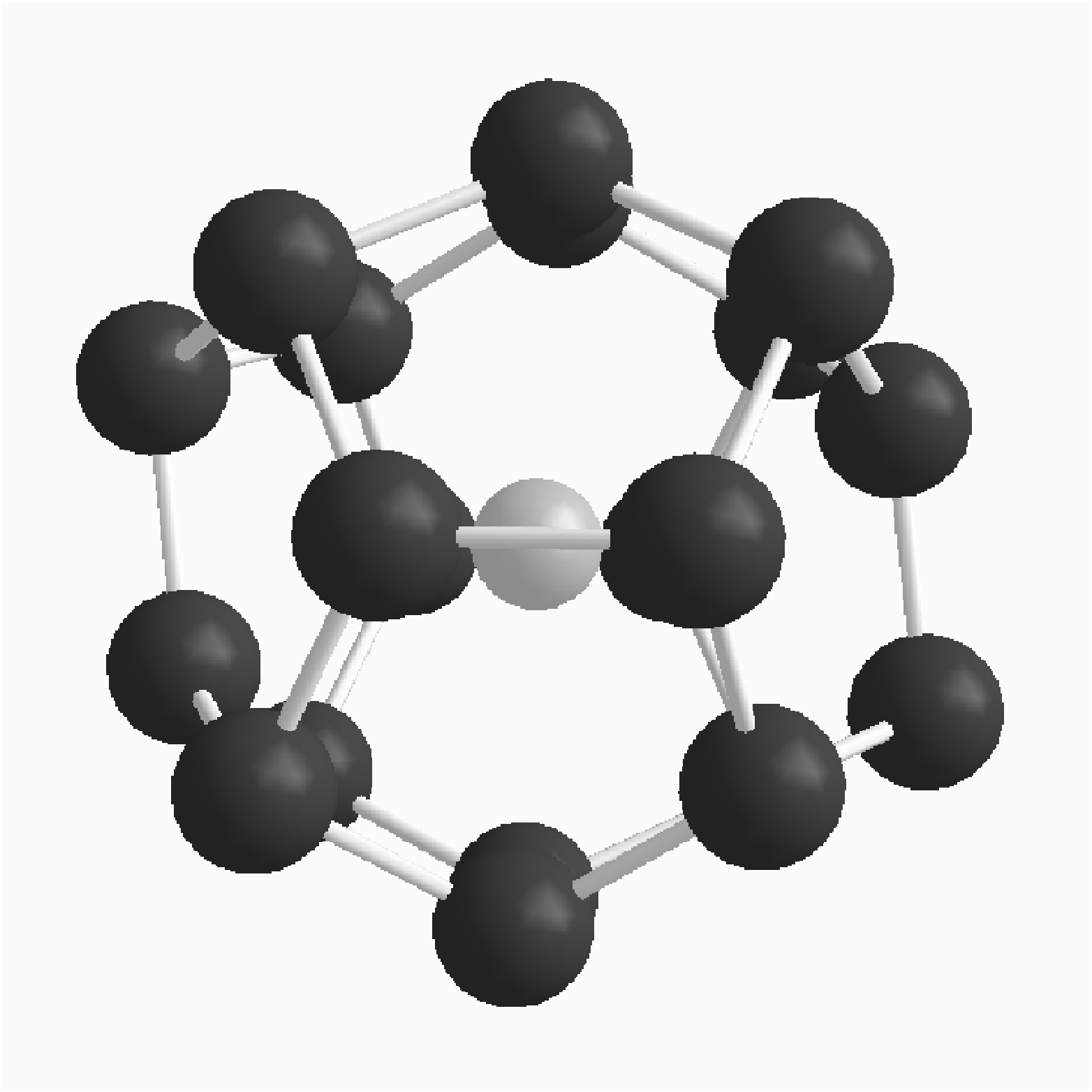}}
\put(9.9,11.1){\includegraphics[width=3.7cm]{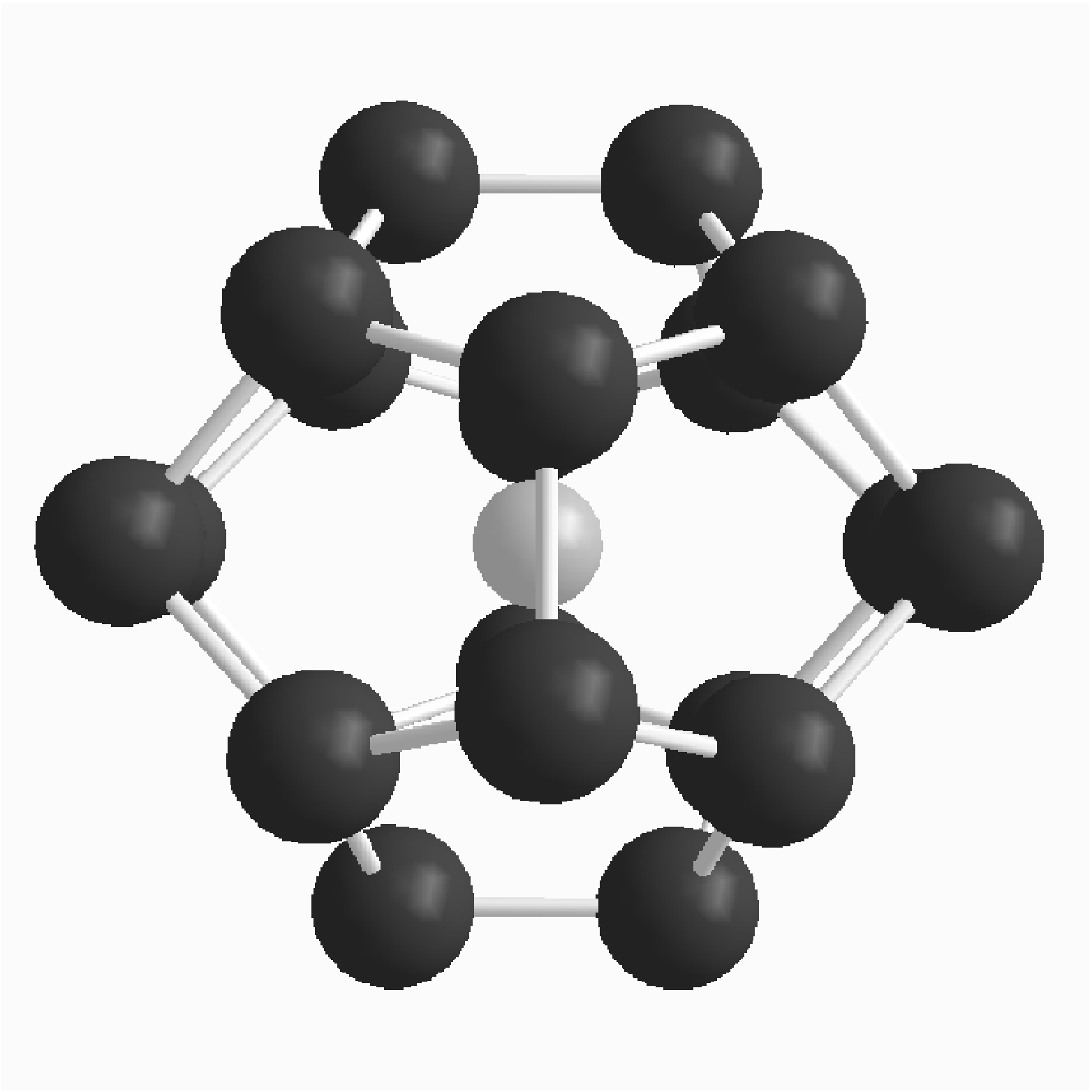}}
\put(9.9,07.4){\includegraphics[width=3.7cm]{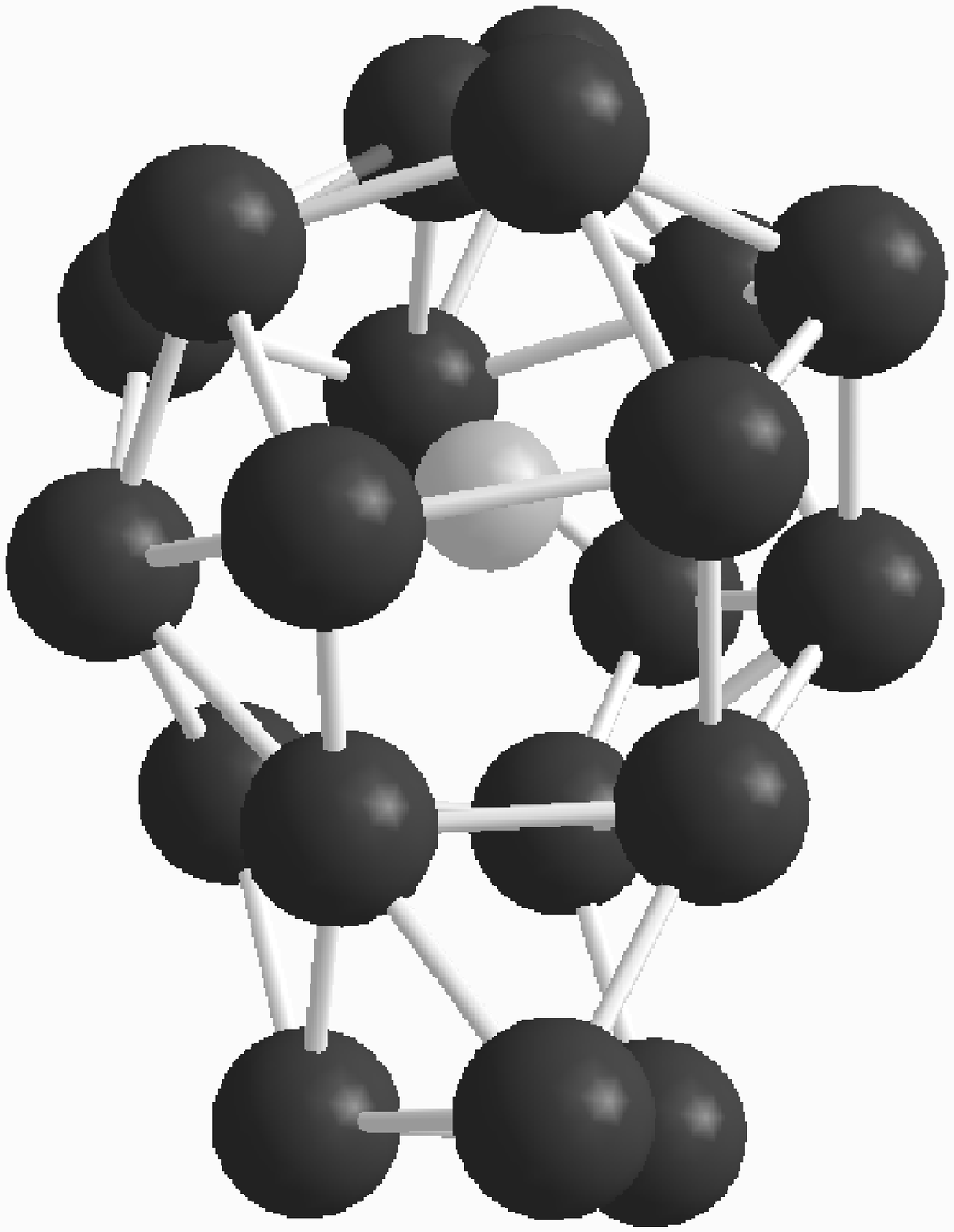}}
\put(9.9,03.7){\includegraphics[width=3.7cm]{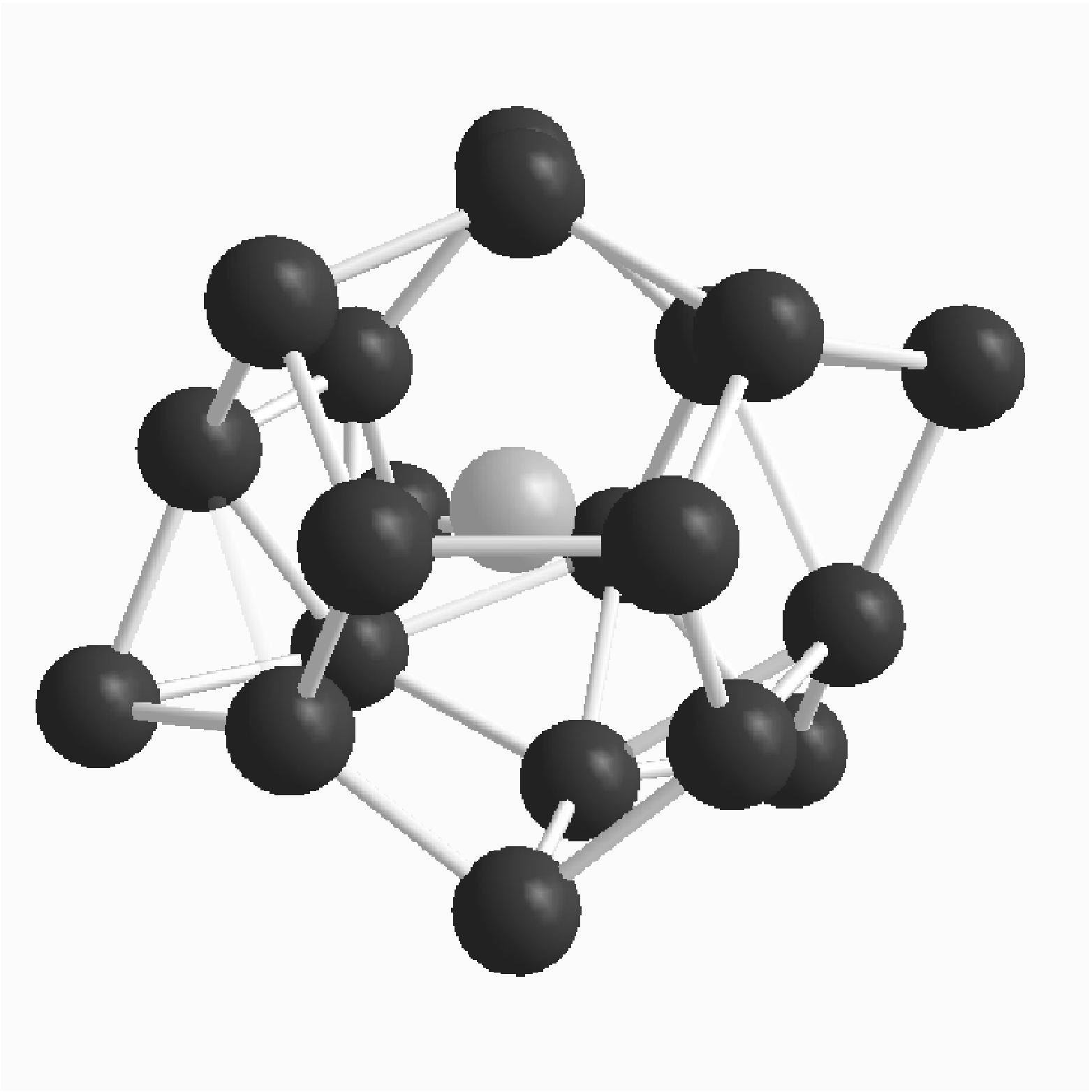}}
\put(9.9,00.0){\includegraphics[width=3.7cm]{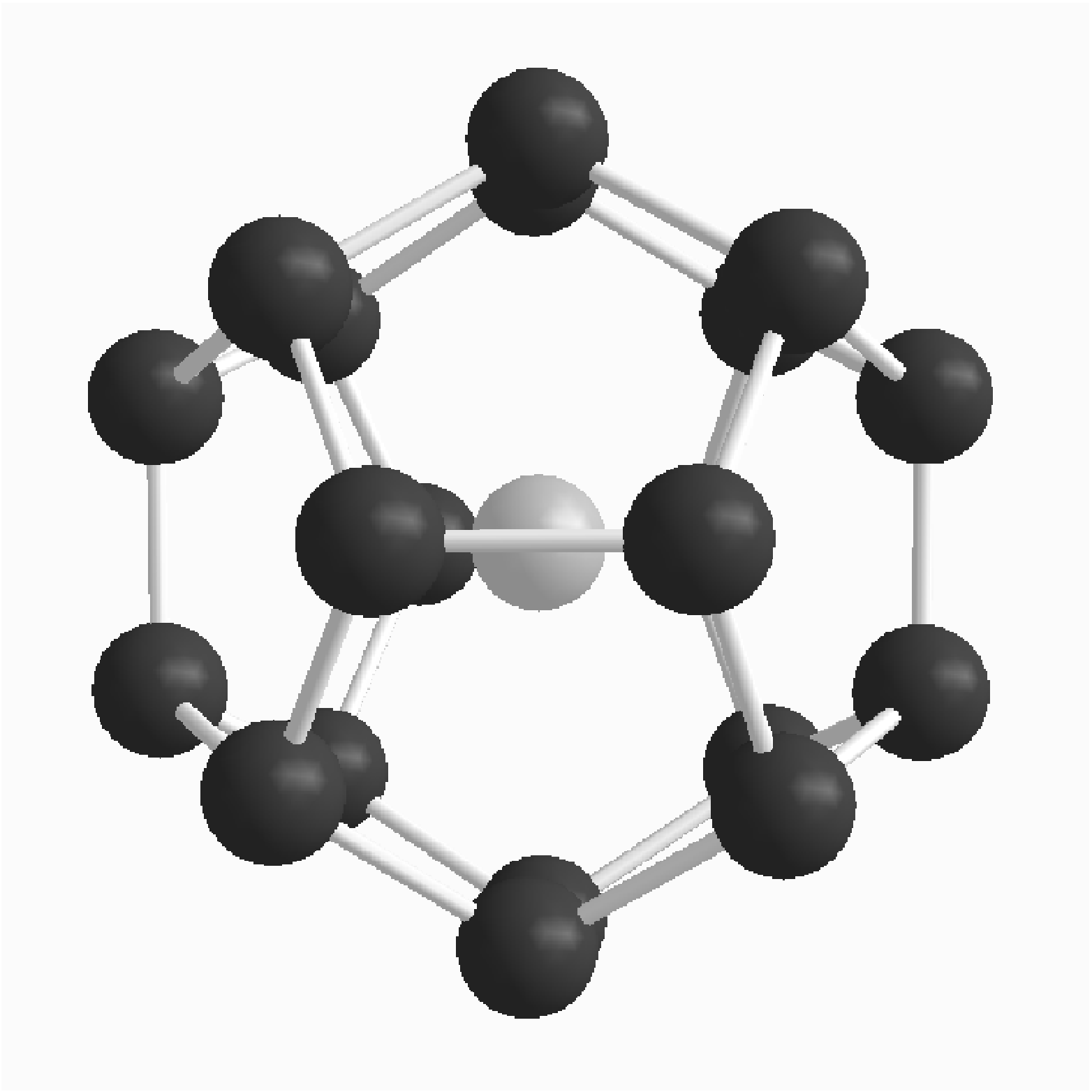}}
%fourth coloumn
\put(13.9,18.5){\includegraphics[width=3.7cm]{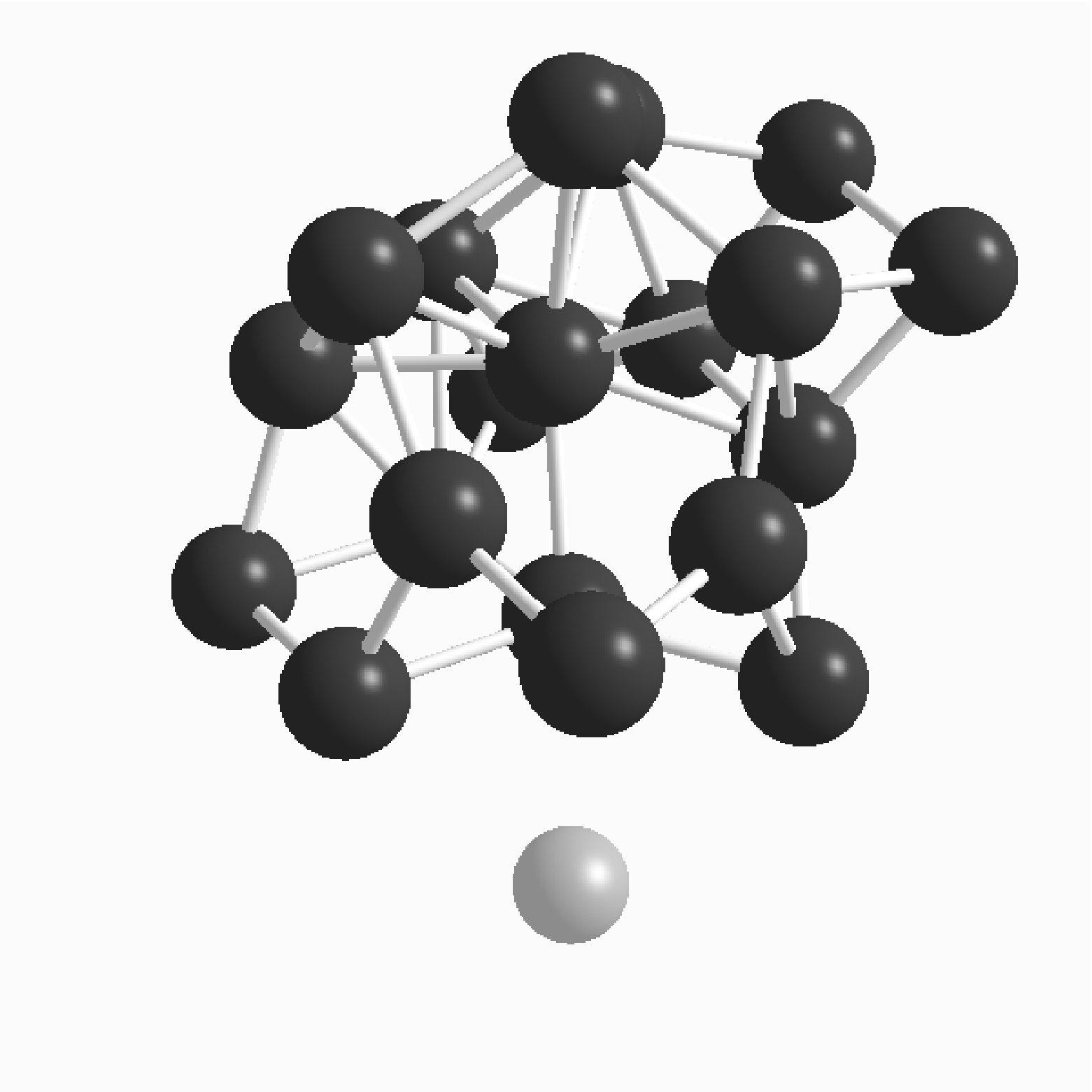}}
\put(13.9,14.8){\includegraphics[width=3.7cm]{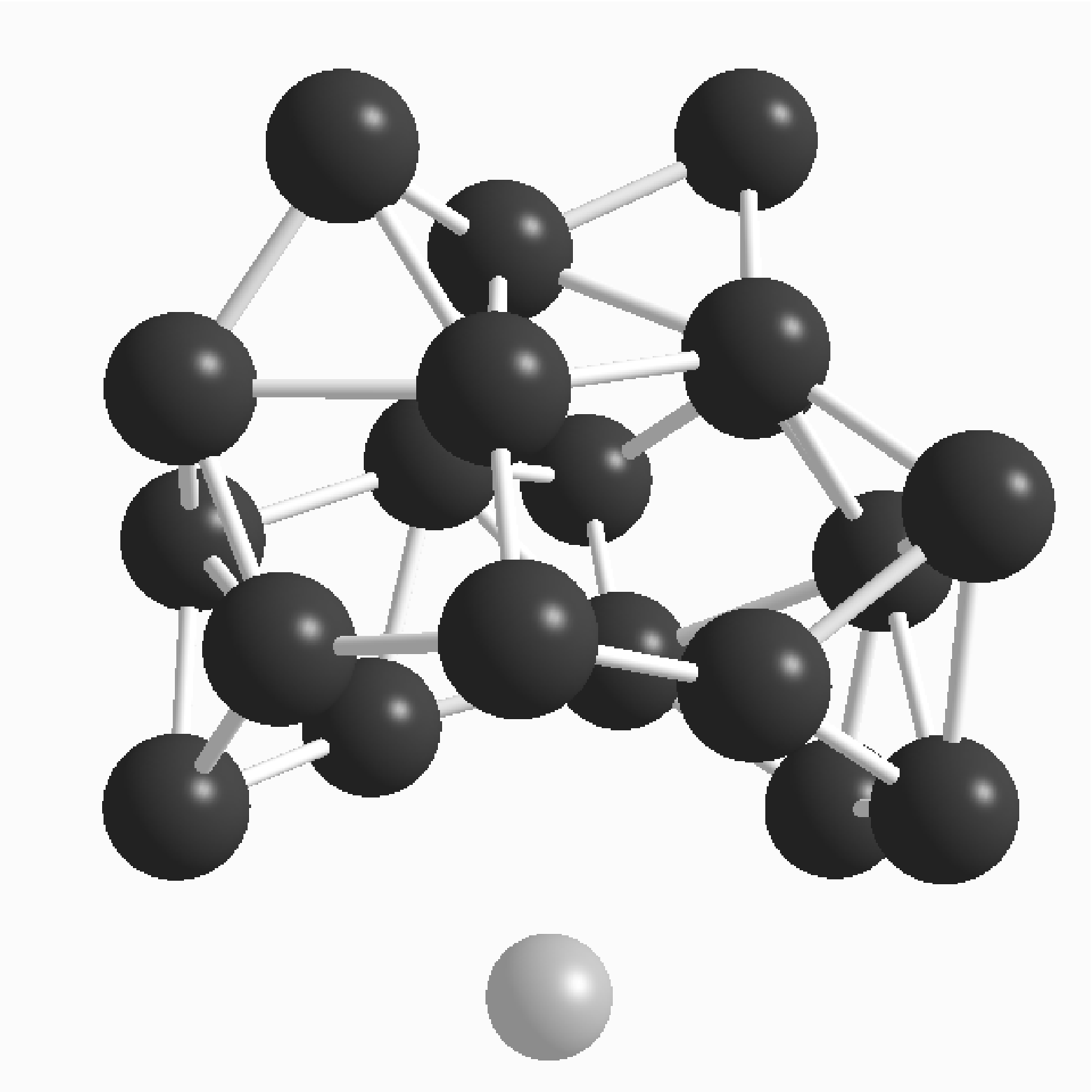}}
\put(13.9,11.1){\includegraphics[width=3.7cm]{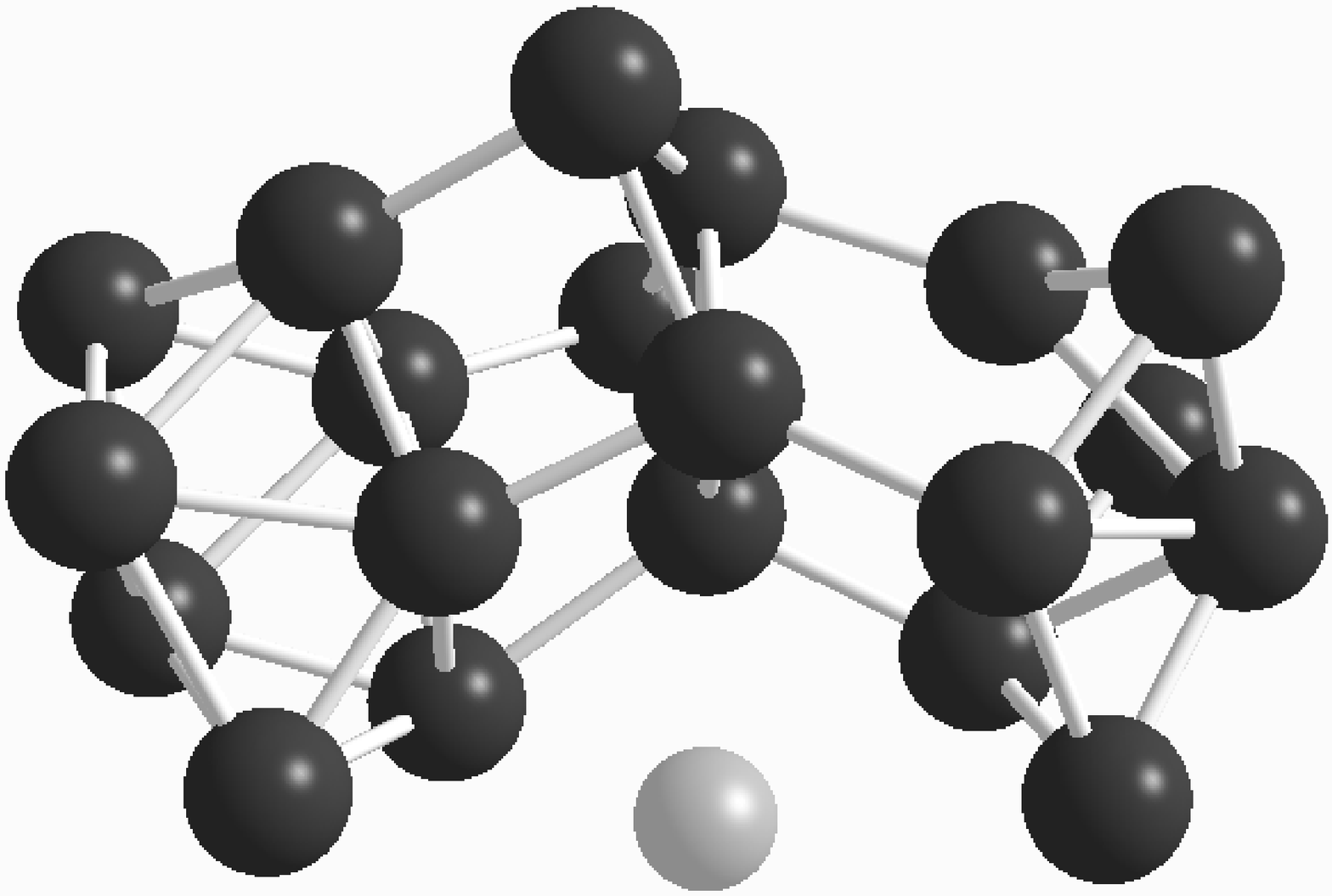}}
\put(13.9,07.4){\includegraphics[width=3.7cm]{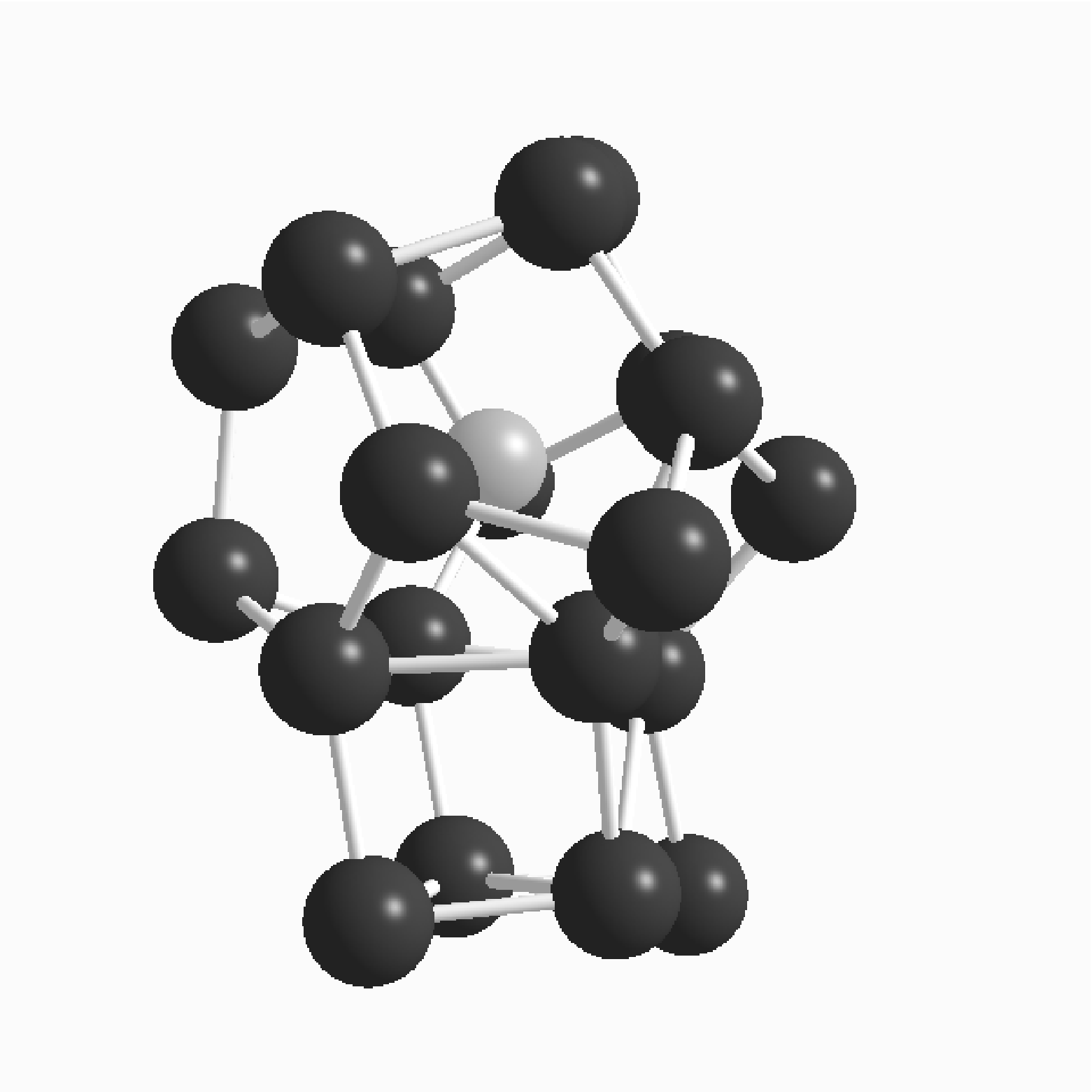}}
\put(13.9,03.7){\includegraphics[width=3.7cm]{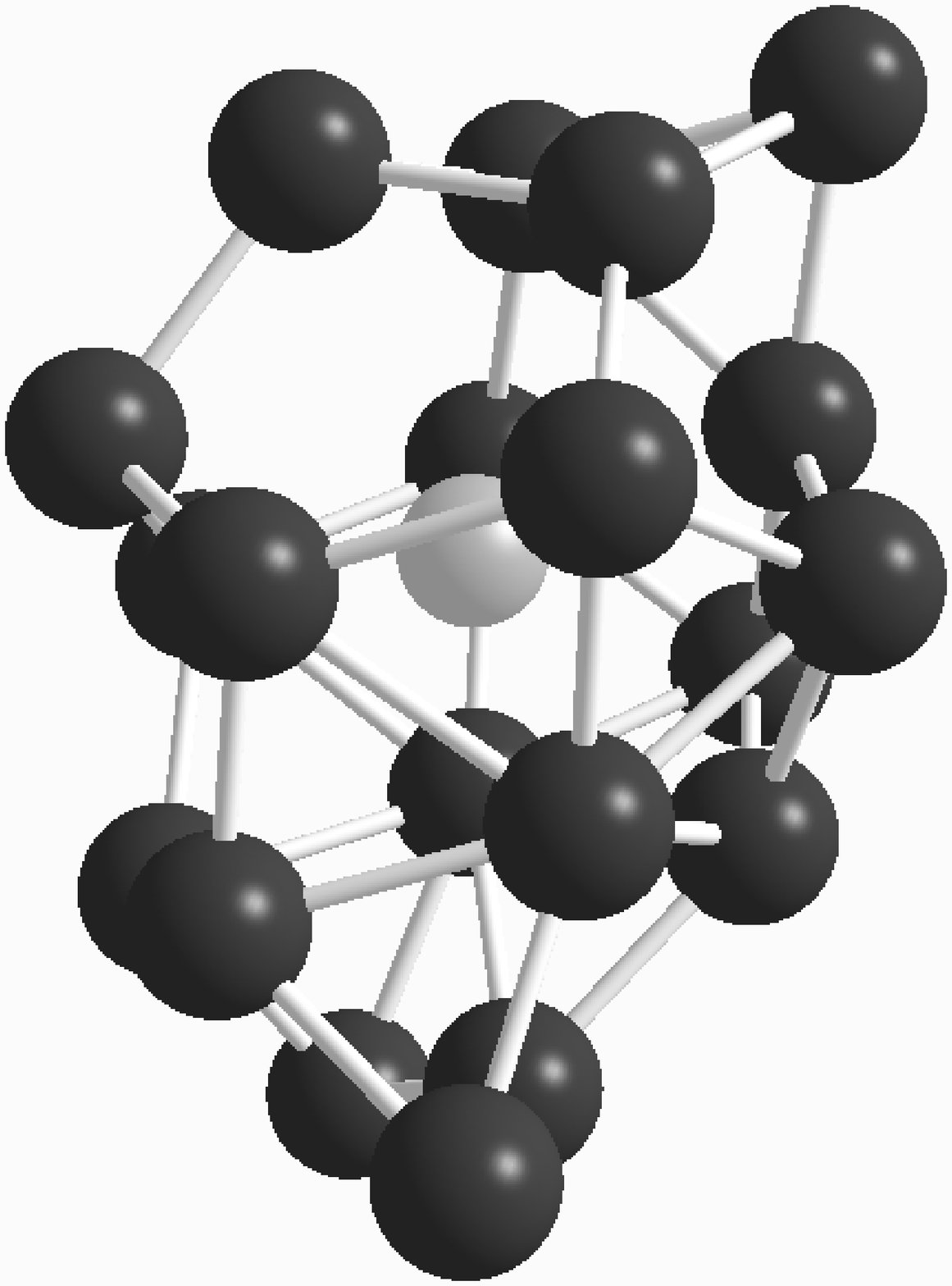}}
\put(13.9,00.0){\includegraphics[width=3.7cm]{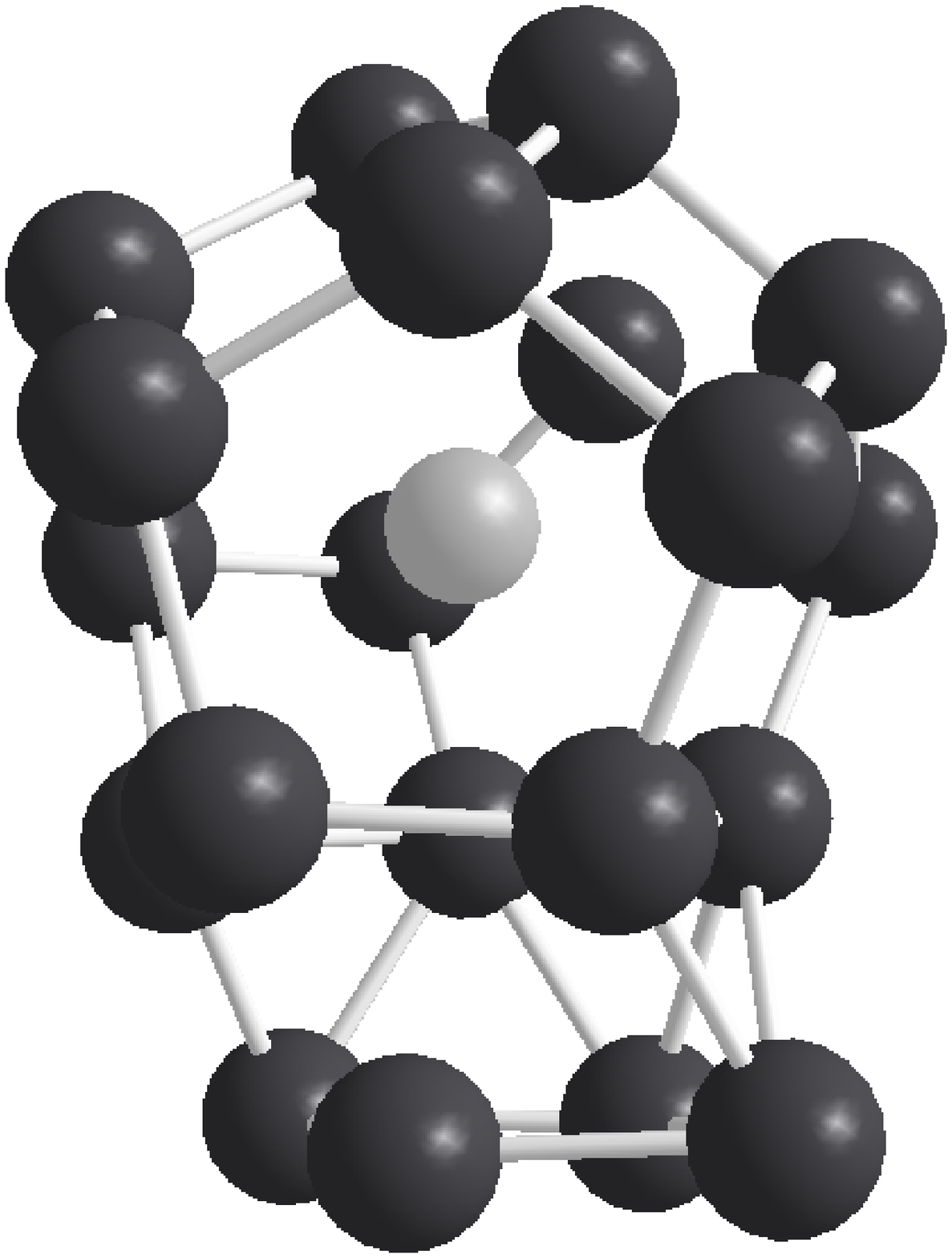}}

\end{picture}
\caption{\label{fig1}
The energetically lowest-lying configurations found for all $M$@Si$_{20}$ systems as shown in the second and fourth columns have either lost their cage structure and dopant (bright) or undergone a transition to a smaller distorted cage with some peripheral silicon atoms. The corresponding most stable (distorted) cage-like structures are given in columns one and three. From top to bottom: $M$ = Ba, Ca, Cr, Cu, K, and Na on the left, and $M$ = Pb, Rb, Sr, Ti, V and Zr on the right half.}
\end{center}
\end{figure*}

A natural starting point for our investigation of the structural stability of metal-doped Si$_{20}$ fullerenes are the perfect cages of $I_h$ symmetry as illustrated in Fig. \ref{fig1}. As impurity atoms we consider a wide range of simple and transition metals (Ba, Ca, Cr, Cu, K, Na, Pb, Rb, Sr, Ti, V and Zr), which comprises in particular those elements contained in clathrate materials and those previously proposed to stabilize the Si$_{20}$ cage structure \cite{sun02,kumar07}. Surprisingly, only relatively short MHM searches over a few hundred PES minima were necessary {\em for all of these dopants} to reveal significantly more stable structures that deviate in either of two ways qualitatively from an endohedral fullerene configuration: As summarized in Fig. \ref{fig1} and Table \ref{tableI}, for most impurity atoms exohedral structures were readily identified. In all other cases, the dopant was not expulsed, but encapsulated in a smaller cage with the remaining Si atoms forming an apical bud. For Ti, V, Cr and Cu dopants, this endohedral structure of the lowest-energy configuration is fully consistent with the interpretation of Ar physisorption experiments \cite{janssens07}. The specific size of the identified smaller cages is furthermore in line with a preferred stability of the corresponding CuSi$_{10}$, CrSi$_{15}$, and $M$Si$_{16}$ ($M = $\,Ti, V, Zr) clusters as deduced from their abundance in mass spectra or simple electron counting rules \cite{janssens07,xiao01,koyasu05}. 
TiSi$_{16}$ and TiSi$_{16}$ cages have also already been identified as local minima in DFT calculations~\cite{smallcage}. 

The MHM searches were stopped as soon as configurations of significantly lower energy than the initial symmetric $I_h$ cage were identified. During the runs typically also a number of more favorable configurations were visited, in which the cage was (partly heavily) distorted, but could still be considered intact, cf. Fig. \ref{fig1}. For those dopant atoms ultimately leading to completely broken cages, a distinction of these structures from the lowest-energy exohedral ones is rather unambiguous. As shown in Table \ref{tableI} the corresponding energy gap to the most stable of these identified intact cage structures is in all cases quite large. This holds for all of the employed xc functionals, even though quite some quantitative scatter can be discerned. With such a clear gap, it is unlikely that an intact cage structure would exist that is even lower in energy and that has been missed in the performed finite MHM searches. Instead, we rather expect that in analogy to pure Si clusters there exists a multitude of further disordered exohedral configurations, which are all extremely close in energy to the here identified most favorable structure \cite{hellmann07}. With exohedral cages being also favorable in terms of entropy, we therefore conclude that the hitherto proposed endohedral Si$_{20}$ fullerenes for the corresponding dopant atoms are only metastable. 

For those impurity atoms ultimately encapsulated by a smaller number of Si atoms, already a mere relaxation of the initial $I_h$ cage resulted in rather heavy distortions as illustrated in Fig. \ref{fig1}. Subsequently sampled configurations exhibited more and more pronounced distortions, spanning a rather continuous range up to the smaller cage lowest-energy structure. In this situation the specification of an energy gap to the lowest-energy intact Si$_{20}$ cage is not well defined and we therefore quote in Table \ref{tableI} the energy difference to the initial relaxed $I_h$ cage. Again, this energy difference is sizable in all employed xc functionals. From this and the observed range of increasingly distorted cages we would therefore also rule out for these dopants that more stable fullerene configurations exist that were not identified in the present MHM searches.

The thus disclosed metastability of the endohedral Si$_{20}$ cages for a wide range of dopant atoms is in distinct contrast 
to the high-symmetry carbon fullerenes \cite{dresselhaus96}. The latter correspond to the global PES minimum with a large 
energy gap to the next lowest-energy structures formed by point defects. 
Still, if the barriers 
surrounding the local PES minima corresponding to the endohedral cages are sufficiently high, kinetic trapping might be 
sufficiently long. 
However, it is well known that clusters of low symmetry have a broad distribution of barrier 
heights~\cite{lowbar} and that 
dynamic processes in such systems involve in general the crossing of several barriers. It is therfore 
unlikely that low symmetry cage minima are surrounded only by high barriers. 
Exploring a dozen saddle points of the CaSi$_{20}$ cage we found barriers that were in most cases higher than 1 eV. 
One barrier leading to an opening of the cage was even only 0.85 eV. According to kinetic rate theory such a 
barrier would correspond to a life time not longer than a few seconds at room temperature. 
Of course, this showcase does not allow to exclude kinetic trapping for all studied dopants in general. 
It neither provides a complete pathway from the cage to the exohedral structure. Even though more detailed 
studies of the dynamics of these clusters would be required to precisely predict their life time, 
our present results suggest that larger barriers than those identified here would be required to 
stabilize the metastable CaSi$_{20}$ cage over time spans relevant for materials applications.

In conclusion, we have used DFT based global geometry optimization to reexamine the proposed stabilization of Si$_{20}$ fullerenes through endohedral metal doping. For a wide range of simple and transition metal dopants this readily reveals that the desirable symmetric cage structure is only metastable. Either exohedral compact configurations or endohedral smaller cages with excess Si atoms forming an apical bud are instead significantly more stable. Regardless of whether local, gradient-corrected or hybrid DFT xc functionals are employed, the resulting energy gap of these lowest-energy configurations to the metastable fullerene cage is in most cases in excess of 1\,eV. 

These findings put severe doubts on the dream of silicon based fullerenes as the building blocks for nano sciences. 
Beyond the specific Si$_{20}$ cages examined here, our study furthermore underscores the importance of a systematic exploration of the configurational space when searching for novel nanoscale materials with predictive-quality theory. 
Financial support from SNF and computing time from CSCS are acknowledged. 
We gratefully acknowledge expert discussions with Volker
Blum regarding FHI-aims.

\end{document}